\documentstyle[sprocl]{article}

\def\Journal#1#2#3#4{{#1} {\bf #2}, #3 (#4)}
\def\Preprint#1#2#3{{#1} #2 (#3)}

\def\be{\begin{equation}}
\def\ee{\end{equation}}
\def\bea{\begin{eqnarray}}
\def\eea{\end{eqnarray}}

\begin{document}

\title{
GRAVITATIONAL LENSING AS A COSMOLOGICAL PROBE
}

\author{ J.-P. KNEIB }

\address{Observatoire Midi-Pyr\'en\'ees, 14 Avenue E. Belin, 31400 Toulouse}


\maketitle

\abstracts{Gravitational Lensing is an efficient tool to probe: the
mass distribution of collapsed systems: galaxies and clusters; high
redshift objects thanks to the gravitational amplification; and the
geometry of the Universe.  I will review here some important
aspects of lensing and related issues in {\it observational}
cosmology.}
  
\section{Introduction}
Geometry of Space-Time depends on its mass content, therefore the
wave-front of cosmologically distant sources, traveling through the
Universe, will be curved by any mass fluctuation en-route.  The more
concentrated the mass fluctuation along the light-of-sight the larger
the distortion of the wave front. For collapsed systems like galaxies
or galaxy clusters the `angular' mass column-density in the central
region is high enough (larger than $\Sigma_{crit}$) to `break' the
wave-front into multiple wave-fronts leading to a {\it multiple number
of images}. The multiple wavefronts, reach us at different times, the
{\it time delay} (td) between images is proportional to angular
distance and hence to the Hubble constant $H_0$.  In the outer parts
of collapsed systems, the `angular' mass column-density is much
smaller (smaller than $\Sigma_{crit}$) and the wave-front is only
slightly distorted, thus the signature of the mass is detected by a
gravitational-shear of the background images.  Only {\it distant} mass
concentrations will act as lenses as the wave-front distortion gets
larger when propagating to larger distances through the Universe.
Lensing is sensitive to the {\bf mass}, {\bf expansion factor} and
{\bf distances}, it is thus a {\it perfect cosmological tool}.\\ This
review is observationally motivated and will focus on two main areas:
{\it Field Lensing} where the targeted gravitational structures are
galaxies and Large-Scale-Structure (LSS), and for which field/all-sky
surveys are the primary source of data; {\it Cluster Lensing} where
the cluster mass distribution can be measured accurately and can then
be used to probe the high-redshift Universe thanks to gravitational
amplification.

\section{FIELD LENSING -- From Strong to Weak}

\subsection{Golden Lenses -- the Perfect Lenses}

{\it Golden lenses} \cite{goldenlens} are those lenses for which
time-delays are or can be measured accurately, and given a 
{\it physically motivated} mass model the value of $H_0$ can be estimated.
Three systems currently deserve this honour \cite{williams97}:
{\bf Q0957} \cite{kundic97a}
($H_0=61$, $\varepsilon_{td}\sim 1\%$, 
$\varepsilon_{model}\sim 10\%$)
{\bf PG1115} \cite{courbin97}
($H_0=53$, $\varepsilon_{td}\sim 10\%$, 
$\varepsilon_{model}\sim 15\%$)
{\bf B0218} \cite{williams97}  ($H_0=70$, $\varepsilon_{td}\sim 25\%$, 
$\varepsilon_{model}\sim 30\%$).
More systems are potentially gilded, but although deriving a 
simple mass model can always be done, the critical part is the measurement
of the time-delay.  Kundic et al \cite{kundic97a} have demonstrated 
that such a precise measurement is feasible, and we can hope for a handful 
of measurements in the years to come, which will then provide
a {\it cosmological} estimate of the Hubble constant.

\subsection{Dark Lenses -- Mysteries ?}

{\it Dark lenses} are lenses with unusually high Mass-to-light ratio (M/L) 
derived either from wide separation images and/or the lack of optical/IR
detection of the stellar component of the primary lens.  Recently 3
{\it dark lenses} have been examined in detail: {\bf MG2016} (the {\it
dark cluster}): Hattori et al \cite{hattori97} have detected using
ASCA/ROSAT a $z_{Fe}\sim 1$ X-ray cluster. However, galaxies seem to
be missing from the cluster compared to its X-ray luminosity;
{\bf Cloverleaf}: Kneib et al \cite{kneib97} have detected an overdensity
of red galaxies within 15" of the quad - suggesting the existence of a
$z\sim 1.7$ cluster, this could help to explain why the primary lens
has not been detected yet; {\bf Q2345+007} (a complex puzzle): Bonnet
et al \cite{bonnet93} reported a weak shear field near the double QSO
more or less centered on Pell\'o et al \cite{pello96} cluster candidate at
$z_{photom}\sim 0.7$, however no rich cluster-like X-ray emission has been
detected with ROSAT and ASCA (Hattori, priv. com.)  in this
field. Fisher et al \cite{fisher94} find clues of $z\sim 1.5$ galaxies
near the double-quasar.  However, adding all these pieces of evidence,
does not produce a simple explanation - is it really a lens system or
a physical pair?

\subsection{Search for New Multiple-Images Systems -- More Lenses !}

Searching for a larger and possibly un-biased sample of multiple image
systems is an important task to increase the sample size
which will allow us to: $\bullet$ better understand the mass distribution
of lenses; $\bullet$ increase the number of possible golden lenses;
$\bullet$ enable statistical analyses of lenses.\\
The JVAS/CLASS \cite{patnaik92}$^{\!,\,}$\cite{myers95} surveys
analysed more than 10000 flat spectrum radio-sources, and find 12
probable lenses, which are currently being studied at higher
resolution in the radio and in the optical wave-bands to identify the
deflector and determine the redshift of the lens and source.
Extension of this survey to wide-separation multiple-images
is currently in progress (see Marlow this conference).
Radio selection is probably the most efficient and un-biased provided we
have a good understanding of the redshift distribution of these
sources \cite{kochanek96}.  An {\it unexpected} result is that  new
lenses discovered have been classified as Spirals or S0s
({\it e.g.} {\bf B1600} \cite{jh97}, {\bf B0218} \cite{hjorth97}). 
This new discovery has motivated the development of a
number of new lensing model for spiral galaxies
\cite{schmidt97}$^{\!,\,}$\cite{perna97}$^{\!,\,}$\cite{kk97b}
(see also Koopmans this conference).

\subsection{Deep Imaging and Spectroscopy -- Unveiling the Complexity}

As underlined above, studying the environment of lenses can be of
great importance in mass modelling. Both deep multi-wavelength
(X-ray, optical, IR)  high-resolution
imaging ({\it e.g.} NIC2/WFPC2 Falco dataset) and spectroscopy 
are therefore of prime
importance. Spectroscopy of the surrounding galaxies has been
undertaken, and surprisingly, in 3 cases a group of galaxies has been
detected: {\bf PG1115} \cite{tonry97}$^{\!,\,}$\cite{kundic97b}
$N_{gal}=5$, $\overline{z}\sim 0.33$, $\sigma_{los}\sim 300$ km/s;
{\bf B1422} \cite{tonry97} $N_{gal}=6$, $\overline{z}\sim 0.34$,
$\sigma_{los}\sim 600$ km/s; {\bf HST14176} \cite{koo96} $N_{gal}=10$,
$\overline{z}\sim 0.81$, $\sigma_{los}\sim 450$ km/s.  Similar
spectroscopic investigations are currently underway in other systems,
namely {\bf MG2016}, {\bf Q2345} and {\bf Q0957}.

\subsection{Mass Distribution -- Toward the Solution}

Keeton, Kochanek \& Seljak \cite{kks97} show that quads are best
modeled by the sum of an elliptical Singular Isothermal Sphere (SIS)
mass model and an additional component of pure external
shear. However, in some cases there is a discrepancy between the
orientation of the light and the mass.  Besides, the magnitude of the
modeled external shear is too large to arise from LSS fluctuations,
but it is likely the results of \cite{hk98}: $\bullet$ the primary
lens (if $\rho$ is steeper [resp. flatter] than a SIS then the shear
is smaller [resp. larger] than a SIS model), however the change of
slope can not easily account for the orientation discrepancy;
$\bullet$ nearby galaxies in a group or a cluster, can easily
account for the external shear and the orientation discrepancy.\\ The
number of constraints for a simple multiple-image system goes like
$C(N-1)+L$, where C is the number of parameter per image (C=3 for a
quasar), N the number of multiple images (N=4 for a quad, N=2 for a
double), L the number of external constraints on the lens - the
stellar component - (center position, ellipticity, orientation,
profile).  Obviously, quads are much preferred specially when the
stellar component is observed (L large).  It is important to increase
the number of free parameters - in particular the mass profile of
lenses - to match the high quality of data now available in optical
(HST) and in radio (MERLIN), however the model solution has to remain
{\it physically} motivated and should include the stellar
component. Saha \& Williams \cite{saha97} recently proposed an
alternative non-parametric modeling based on linear programming, where
the mass distribution is `pixelised'. Although, it does indeed
increase the number of parameters, the current technique suffers from
the fact it allows too many degrees of freedom, compared to the
available number of constraints.

\subsection{The Case of HST14176}

The {\bf HST14176} \cite{rat95} multiple system has been found among
HST/WFPC2 survey of field galaxies - where we expect roughly 1 such
system within $\sim$100 arcmin$^2$.
The lens is a $z=0.81$ \cite{crampton96} elliptical galaxy (M$_B$=-22.9), 
the source is at $z=3.4$ \cite{crampton96}.  Hjorth \& Kneib  \cite{hk98}
modelling consistently the
stellar and dark component, as well as surrounding galaxies,
are able for the first time to put an upper limit 
(maximum light hypothesis) on the stellar M$^*$/L$_B<2.5 h_{50}$ 
and on the expected stellar line-of-sight
velocity dispersion $\sigma_{los}\,<270$km/s.
Folding this result into the Fundamental Plane \cite{jorgensen96},
allows the computation of an
independent measure of the evolution of elliptical galaxies, which is
compatible with synthetic evolution model of elliptical galaxies.
That particular example demonstrates that detailed modeling of 
a particular system can be of great interest.

\subsection{Galaxy-Galaxy Lensing -- Mass Profile at Large Radius}

Distant galaxies not aligned with the lensing galaxy, will not be
multiply imaged but will suffer a small distortion by the lens.  The
high density of faint galaxies allows us then to probe in a
statistical way the mass, and mass profile of an `average' galaxy to
larger distance ($\sim$100$h_{50}^{-1}$ kpc) compared to multiple
images which mainly concentrate on the $\sim $1 to $\sim
$20$h_{50}^{-1}$ kpc range.  It is only recently that observations
have been successful
\cite{bbs96}$^{\!,\,}$\cite{griffith96}$^{\!,\,}$\cite{dat96} to
detect this effect.  The main limitation in this technique is the
errors in the determination of shape parameters of faint (distant)
galaxies, the unknown redshift distribution of galaxies, the adopted
scaling laws and finally the limited size of the sample.  Schneider \&
Rix \cite{schneider96} proposed a maximum likelihood method to
consistently take into account the effect of multi-plane and
multi-deflector lensing.  An improvement of this method (that takes
into account detection noise) is in progress by Ebbels, Kneib \& Ellis
\cite{ebbels98} and is applied to a large HST/WFPC2 archival dataset,
which includes $\sim 250$ measured redshifts for the brightest
galaxies.  The mass derived is consistent with dynamical estimates in
the central part, and requires extended dark halo, although its extent
is currently not well constrained.

\subsection{Large Scale Structure -- Yet to Be Detected}
Although, we do expect weak gravitational shear from LSS
\cite{blandford91}$^{\!,\,}$\cite{miralda91}$^{\!,\,}$\cite{kaiser92}$^{\!,\,}$\cite{kaiser96},
the expected signal on degree scale is of the order of 2 to 5\%
depending on the cosmology. Its small amplitude have precluded up to
now any secure and positive detection \cite{mould94}.  The
no-detection is in turn triggering a great deal of efforts: $\bullet$
to find the best estimator that can constrain both the power spectrum
and the cosmological parameters \cite{bernardeau97}; $\bullet$ to
determine the best method to detect/correct the weak shear signal
\cite{ludo97}; $\bullet$ and to develop efficient wide-field
mosaic-CCD cameras.\\ Two strategies can be followed to measure LSS
weak shear signal: $\bullet$ observe a wide contiguous field (of the
order of a few square degrees), and extract the power-spectrum of mass
fluctuations, from small to large scales; $\bullet$ observe random
fields {\it deep and wide} enough to beat the noise of the width of
the ellipticity distribution and allow a significant measure of a mean
gravitational shear on each individual field. The mean shear
statistics in the various fields can then be folded to constrain the
mass power-spectrum at the observed scale.

\section{CLUSTER LENSING -- The New Area}

\subsection{Multiple-Images Modelling -- Understanding Cluster Cores}

The discovery  in the late 80's of giant arcs \cite{soucail87}
(results of the merging of faint galaxy multiple images on a critical line)
in cluster cores opened a new window for lensing investigation. 
It was quickly understood that multiple-images are 
{\it key constraints} on the central mass distribution of cluster-lenses
\cite{kkf92}$^{\!,\,}$\cite{mellier93}$^{\!,\,}$\cite{kneib93}, furthermore
if the redshift of one giant-arc/multiple image is known then 
the mass can be  calibrated in an absolute way.
Therefore, the redshift of any other multiple-images can then 
be estimated \cite{kneib93}$^{\!,\,}$\cite{kneib95},
as their angular separation is a function of their redshift.
The mere existence of these gravitational arcs lead to the demonstration
that cluster total mass profile require a small `core radius'
\cite{kneib93}$^{\!,\,}$\cite{smail96}.\\
The first cluster-lens images acquired with the HST/WFPC2 camera
\cite{kneib96}$^{\!,\,}$\cite{colley96}$^{\!,\,}$\cite{seitz97}
unveiled the complex nature of the mass distribution in the cluster
core: indeed any bright (typically L$^*$ or brighter) galaxies
contribute importantly to lensing effects
by increasing the multiplicity of some images and 
slightly changing the location of others. The large number 
of multiple-images
revealed by HST (thanks to its high-resolution)
allow us to reconstruct a detailed mass
distribution that include not only the cluster mass but also
cluster-galaxy halos. The detailed theoretical
interpretation was first  presented
by Natarajan \& Kneib \cite{nk97}, who futhermore show that the
galaxy-galaxy lensing approach can be extended to cluster-field.\\
The converging picture of cluster-core is a global M/L$_V$($<300
h_{50}^{-1}$)=140$\pm20 h_{50}$ for the cluster and
M/L$_V$=12$\pm5 h_{50}$ for galaxy halos \cite{nkes97} (see
also Natarajan - this conference for a more extended discussion).
These results compare nicely with the recent high-resolution
numerical simulations of galaxy clusters 
({\it e.g.} Moore et al this conference).

\subsection{Cluster Weak Lensing -- mass profile at large radii}

It is now well known that weak gravitational shear of clusters can be
inverted to obtain mass estimates of
clusters \cite{ks93}$^{\!,\,}$\cite{ss95} and a wealth of
observations have detected the signal in many different clusters
{\it e.g.} {\bf Cl0024} \cite{bonnet94}, {\bf MS1224} \cite{fahlman94},
{\bf A2218} \cite{squires96}, {\bf A2390} \cite{squires96b},
{\bf A2163} \cite{squires97}, {\bf A1689}  \cite{fisher97},
{\bf Cl0939} \cite{skss96} - (see also Van Kampen et al this conference).\\  
The requirements for
successfully using weak lensing to constrain the mass distribution are
the following: $\bullet$ high resolution images (HST or sub-arcsec
seeing ground based images); $\bullet$ wide field camera; $\bullet$
deep images in order to increase the number density of faint galaxies;
$\bullet$ and a {\it good} understanding of the ellipticity correction
that would otherwise easily induce an under- or over-estimate
of the total mass.\\
The main results are: $\bullet$ mass profile is compatible with
$\rho\sim 1/r^2$;  $\bullet$ M/L($<1 h_{50}^{-1}$ Mpc) $\sim 200
h_{50}$, but in some cases one gets M/L $\sim $ 300-400. \\
The new developments are: 
$\bullet$ to probe mass of galaxy groups
$\bullet$ to probe the central mass profile in low-$z$
clusters \cite{luis98}; $\bullet$ to probe structures between medium-$z$
clusters \cite{luppino98}; $\bullet$ to weight high-$z$
clusters \cite{smaildick95}$^{\!,\,}$\cite{luppino97}
to constrain cluster evolution and the mean redshift-distribution of faint
galaxies.

\subsection{Comparison with Other Mass Estimates: X-ray and Dynamics}

Miralda \& Babul \cite{miralda95} noticed that lensing and X-ray estimate
within the arc-radius differs significantly by a factor of $\sim 2$.
This discrepancy can however easily be resolved when proper
careful lensing and X-ray analysis are performed. Allen, Fabian \& Kneib
\cite{allen95} have shown that in the cooling-flow cluster {\bf PKS0745},
multi-phase X-ray model does agree with the lensing mass estimate (but
a simple X-ray mass model would be discrepant by a factor of 2--3).
Allen \cite{allen98} has further extended this comparison to a larger 
set of clusters. X-ray cooling flow clusters with a multi-phase model
do agree with lensing mass estimates. For non-cooling flow clusters
however, one has to require the total mass to have a smaller
core radius than one expected from the X-ray data to match the lensing
mass estimates. Clearly, a systematic and detailed study will enable
a better understanding of the ICM and its dynamical state.\\
HST observations \cite{smail97} are and will be
(specially with the forthcoming Advanced Camera for Survey)
ideally suited  to study the physics of galaxy clusters when combined
with the upcoming generation of X-ray telescopes (AXAF/XMM).\\
Natarajan \& Kneib \cite{nk96} have shown that lensing can help
to understand the dynamics of galaxies in the global potential well.
The velocity anisotropy can be recovered showing evidence for the
nature of orbits in the virialized core, as well as orbits in the
outskirts \cite{tormen97}.

\subsection{Cluster as Low Resolution Spectrograph -- 
Distance of Faint Galaxies}

The probability redshift distribution of an arclet for a given mass of
the cluster and the shape of the arclet depends only on the
ellipticity distribution of faint galaxies \cite{kneib96}: the larger
the redshift, the larger (in general) the deformation induced by the
cluster. From a secure mass distribution defined by a few sets of
multiple-images, we can then estimate the likely $z$ of arclets behind
well constrained clusters.  This is most interesting as this method is
purely geometrical and therefore does not suffer from the
spectroscopic bias that provides redshifts only for those faint
galaxies with (strong) optical emission lines.  To evaluate the
accuracy of this technique on a cluster-lens {\bf A2218}
\cite{kneib96} Ebbels et al \cite{ebbels97} have successfully measured
on WHT the redshift of 19 arclets providing a first confirmation of
the lensing method. Similar work is now in progress in other cluster
lenses such as {\bf A2390} \cite{kneib98} and {\bf AC114}.

\subsection{Statistical Analysis}

The number of arcs/arclets behind cluster cores is a complex
function of the detailed mass distribution, and the evolution
of faint galaxies.\\
On a individual cluster-basis, if the mass distribution is well understood,
(granularity in the mass distribution can dramatically change
counts of arclets, in a redshift dependent way)
the statistics of arclets can be a possible way to constrain
galaxy evolution \cite{bezecourt97}, and in particular puts limits 
on the UV emission of distant galaxies - providing we understand
correctly the surface-brightness and detection biases.\\
Using a cluster sample, the number of arcs/arclets will be
sensitive to cluster masses (therefore on cluster evolution
which critically depends on the density parameter $\Omega$)
and the volume element ({\it i.e.} the cosmological constant $\lambda$)
(see Bartelmann \& Bahcall this conference for a more complete discussion).
Therefore, counting arcs in clusters may constrain models of cluster 
evolution and therefore the values of cosmological parameters.

\subsection{Probing the Distant Universe -- New Windows}

Cluster-lenses offer a large amplification (typically 1 to 2
mag) over a few arcmin$^2$.  Hence these clusters can be used as
{\it telescopes} to probe the distant Universe, and detect sources
that would otherwise be extremely difficult to detect.  A particular
application is the detection of `normal' distant $z>2$ galaxies behind
clusters: $z=2.24$ (Cl2244) \cite{mellier91}, $2.55$ (A2218) \cite{ebbels96}, 
$2.72$ (MS1512) \cite{yee96},
$3.33$ (Cl0939) \cite{trager97}, $3.98$ (Cl0939) \cite{trager97},
$4.05$ (A2390) \cite{pello98}, 
$4.92$ (Cl1358) \cite{franx97}.  An other
promising avenue 
is the first identification in yet unexplored
windows like IR/sub-mm/mm bands of distant $z>3$ objects - because we
do expect to detect redshifted dust-emission from these
distant galaxies \cite{blain97}.  A first tentative application in
sub-mm has been recently presented by Smail, Ivison \& Blain\cite{sib97}.

\section{Future and Prospects}

Thanks to improvement in lensing techniques and deeper, wider,
higher-resolution images, {\it lensing} is rapidly becoming a useful
{\it tool} to weigh the Universe (from small to large scales), probe
the distant Universe and constrain the evolution and its fundamental
parameters.\\ In the years to come we might have: a lensing estimate
of $H_0$ with 2 significant digits; a lensing estimate of $\Omega$
with 1 significant digit and an upper limit on $\Lambda$; a good
knowledge of the mass distribution of collapsed systems (galaxies and
clusters), and therefore the physics of the ICM as well as the mass
exchange between cluster and galaxies; some constraints on the mass
power spectrum; a better understanding of the redshift distribution of
faint galaxies and their evolution; a better view of the distant
Universe and its dusty galaxies.

\section*{Acknowledgments}
{
Thanks to the Conference organisers and most specially
Joachim Wambsganss and Volker M\"uller. 
Thanks to Yannick, Richard, Ian, Priya, Timothy, Jens, Makoto,
Peter, and the Toulouse Team for the many discussions we 
had on the many beautiful
lenses and the many mysterious objects behind. 
I acknowledge support from the organizing committee as well as CNES/INSU.
}

\section*{References}
{

}

\end{document}